\documentclass[onecollarge,fleqn,runningheads]{svjour2}
\smartqed  
\usepackage{graphicx}
\usepackage{xcolor}
\usepackage{mathptmx}      
\usepackage{psfrag}
\usepackage{amsmath}
\usepackage{amssymb}
\usepackage{array}

\newcommand{\ie}{\emph{i.e.~}}

\newcommand{\bm}[1]{{\mathbf #1}}
\newcommand\Rey{\mbox{\textit{Re}}}  
\newcommand\Ubase{{\bm{U}}}
\newcommand\Pbase{P}
\newcommand\upert{{\bm{u}'}}
\newcommand\ppert{{p}'}
\newcommand\umode{\tilde{\bm{u}}}

\newcommand\NavSto{Navier--Stokes}
\newcommand\NS{\NavSto}
\newcommand\LNS{linearized \NS}

\newcommand\xvec{\bm{x}}
\newcommand\uvec{\bm{u}}
\newcommand\uadj{\bm{u}^*}

\newcommand\padj{p^*}
\newcommand\cd{\mathrm{d}}
\newcommand\Aop{{\cal A}}
\newcommand\Aadj{{\cal A^*}}

\newcommand\twod{two-di\-men\-sion\-al}

\newcommand\opt{\textrm{max}}
\newcommand\Gmax{G_\opt}
\newcommand\U{U_\infty}

\journalname{Theoretical and Computational Fluid Dynamics}

\begin{document}

\title{Transient growth mechanisms of low Reynolds number flow over a low pressure turbine blade
}

\titlerunning{Transient Growth over an LPT blade}        

\author{A S Sharma		\and
		N Abdessemed	\and
		S J Sherwin		\and
		V Theofilis
}


\institute{A S Sharma \at
Department of Electronic and Electrical Engineering\\
Department of Aeronautics\\
Imperial College\\
Exhibition Road\\
London\\
SW7 2AZ\\
UK\\
\email{ati@imperial.ac.uk}           
\and
N Abdessemed \at
Transsolar Energie Technik GmbH\\
Curiestra\ss{}e 2\\
D-70563 Stuttgart\\
Germany\\
\email{nadir.abdessemed@gmail.com}           
\and
S J Sherwin \at
Department of Aeronautics\\
Imperial College\\
Exhibition Road\\
London\\
SW7 2AZ\\ UK\\
\email{s.sherwin@imperial.ac.uk}           
\and
V Theofilis \at
School of Aeronautics\\
Universidad Politecnica de Madrid\\
Pza. Cardenal Cisneros 3\\
E28040 Madrid\\
Spain\\
\email{vassilios.theofilis@upm.es}	
}

\date{Received: date / Accepted: date}

\maketitle

\begin{abstract}

A direct transient growth analysis for two dimensional, three component 
perturbations to flow past a periodic array of T-106/300 low pressure turbine 
fan blades is presented. The methodology is based on a singular value 
decomposition of the flow evolution operator, linearised about a steady or 
periodic base flow. This analysis yields the optimal growth modes.
 Previous work on global mode stability analysis of this flow geometry 
showed the flow is asymptotically stable, indicating a non-modal 
explanation of transition may be more appropriate. The present work extends 
previous investigations into the transient growth around a steady base 
flow, to higher Reynolds numbers and periodic base flows. It is found that 
the notable transient growth of the optimal modes suggests a plausible 
route to transition in comparison to modal growth for this configuration.  
The spatial extent and localisation of the optimal modes is examined and 
possible physical triggering mechanisms are discussed. It is found that for 
longer times and longer spanwise wavelengths, a separation in the shear 
layer excites the wake mode. For shorter times and spanwise wavelengths, 
smaller growth associated with excitation of the near wake are observed.

\PACS{First \and Second \and More} 
\end{abstract}

\section{Introduction}
\label{sec:intro}
Hydrodynamic stability is typically studied by the method of linearisation 
and subsequent modal analysis~\cite{DrazinReid}. This approach considers 
the asymptotic behaviour of small perturbations to a steady or 
time-periodic base flow.  Such asymptotic behaviour is determined by the 
eigenvalues of a linear operator arising from the analysis, describing the 
time-evolution of the eigenmodes.

Many canonical problems, such as flow in a channel, permit such stability 
analysis to be performed about a velocity field which depends on a single 
coordinate. However in more complex geometries we can extend the classical 
hydrodynamic stability analysis to use fully resolved computational 
stability analysis of the flow field \cite{BarkleyHenderson,Tuckerman}.  
This is referred to as {\em biglobal stability analysis} 
\cite{TheofilisPrAeS2003} or {\em direct linear stability analysis} in 
analogy to direct numerical simulation (DNS). This approach is able to 
resolve fully the base flow in two or three dimensions and to perform a 
stability analysis with respect to perturbations in two or three 
dimensions. This methodology does not need to resort to any approximations 
beyond the initial linearisation and the imposition of inflow and outflow 
conditions.

In particular the biglobal stability analysis method allows us to consider 
flows with rapid streamwise variation in two spatial dimensions such as the 
case of interest, flow over a low pressure turbine blade.  By postulating 
spanwise homogeneous modal instabilities of the form: $\upert(x,y,z,t) = 
\umode(x,y)\exp(i\beta z +\lambda t)$, asymptotic instability analysis 
becomes a large scale eigenvalue problem for the modal shape $\umode$ and 
eigenvalue $\lambda$. This permits use of algorithms and numerical 
techniques which provide the leading eigenvalues and eigenmodes for the 
resulting large problems, typically through iterative techniques such as 
the Arnoldi method~\cite{Tuckerman}. This approach is extremely effective 
at determining absolute instabilities in many complex geometry flows, both 
open and 
closed~\cite{BarkleyHenderson,EhrensteinPF1996,DingKawaharaJCP1998,hmb02a,TheofilisFedorovObristDallmannJFM2003,TheofilisDuckOwen,shbl05,GonzalezTheofilisGomezBlanco,blsh07,TheofilisHeinDallmann} 
including weakly nonlinear stability \cite{Tuckerman,HendersonBarkley}.

Direct linear stability analysis has not been routinely applied to 
convective instabilities that commonly arise in open domain problems with 
inflow and outflow conditions.  One reason is that such flows are not 
typically dominated by modal behaviour, but rather by significant growth of 
transients that can arise owing to the non-normality of the eigenmodes.  A 
large-scale eigenvalue analysis is not designed to detect such behaviour, 
although for streamwise-periodic flow, it is possible to analyse convective 
instability through direct linear stability analysis~\cite{sbs95}.

To examine this situation, hydrodynamic stability analysis has been 
extended to cover non-modal stability analysis or transient growth 
analysis~\cite{ButlerFarrell,Trefethenetal,SchmidHenningson,Schmid07}. This 
approach poses an initial value problem to find the linear growth of 
infinitesimal perturbations over a prescribed finite time interval. Much of 
the initial focus in this area has been on large linear transient 
amplification and the relationship of this to subcritical transition to 
turbulence in plane shear flows~\cite{farrell88,ButlerFarrell}.

This approach was recently employed in backward-facing-step 
flows~\cite{msj06,blbash08}.  With different emphasis from the present 
approach, Ehrenstein \& Gallaire~\cite{ehrenstein05} have directly computed 
modes in boundary-layer flow to analyze transient growth associated with 
convective instability and H{\oe}pffner~\emph{et~al}~\cite{hbh05} 
investigated the transient growth of boundary layer streaks. A comparable 
study to the present work \cite{Abdessemed09}, studied steady and periodic 
flows past a cylinder. A preliminary study along the lines of the current 
work, concentrating on a steady base flow, is described in 
\cite{shabshth06}.

Methods developed for direct linear stability analysis of the \NavSto\ 
equations in general geometries have been previously described in 
detail~\cite{Tuckerman}, and extensively applied 
\cite{BarkleyHenderson,bgh02,shbl05,hmb02a,bllo03a,blsh07,ebs06}. 
Subsequently, large-scale techniques have been extended to the transient 
growth problem. In \cite{bablsh08} a method suitable for such direct 
optimal growth computations for the linearized \NavSto\ equations in 
general geometries was described in detail. The extension to periodic base 
flows was presented in \cite{BlackburnSherwinBarkley07}, for the case of a 
stenotic/constricted pipe flow. The approach has been applied to steady 
flows past a low pressure turbine blades \cite{shabshth06}, a backward 
facing step \cite{blbash08} and a cylinder, \cite{Abdessemed09}, and is the 
method adopted in this study.

Recent work 
\cite{AbdessemedSherwinTheofilis2004,AbdessemedSherwinTheofilis2006,Abdessemed09-1} 
on the flow past the same T-106/300 low-pressure turbine blade (LPT) as 
used in this study, concentrated on a biglobal stability analysis in order 
to understand the instability mechanisms in this class of flows. At a 
Reynolds number of 2000, the base flow displays periodic shedding. The work 
imposed periodic boundary conditions which imply synchronous shedding from 
all blades. Relevant to the current study, the work found that these 
periodic boundary conditions caused the marginally stable flow to go very 
marginally unstable (Floquet multiplier $\mu$ just over $1$) at a Reynolds 
number of 2000 and spanwise wavelengths $L_z$ of approximately $L_z = 3c$ 
to $L_z = 12c$, where $c$ is the projection of the chord length on the 
streamwise axis. This instability is understood to be due to strict 
enforcement of the periodic boundary conditions, which is arguably not 
physical. Relaxing the strict periodicity by using a double-bladed mesh 
resulted in stable eigenmodes ($\mu <1$) for all Reynold numbers explored, 
presumably because small subharmonic effects were sufficient to supress 
synchronous shedding and allow asynchronous shedding. The same geometry is 
analysed here. For our purposes, the flow is best characterised as 
marginally unstable with Floquet multiplier ($\mu=1.00$). This work 
highlights that modal analysis with secondary instabilities does not 
explain transition in this flow.

Abdessemed et al 
\cite{AbdessemedSherwinTheofilis2004,AbdessemedSherwinTheofilis2006,Abdessemed09-1} 
also considered the transient growth problem using a steady base flow 
at a Reynolds number of 895, where significant transient growth up to 
order $10^5$ was observed. The present study extends this to periodic 
base flow at a higher Reynolds number of 2000, where the base flow is 
periodic.

The paper is outlined as follows. In Section \ref{sec:method} we outline 
the direct stability analysis method and the direct transient growth 
analysis needed to determine the peak growth and the associated 
perturbations. In Section \ref{sec:results} we present results for 
transient growth about a periodic base flow, considering in turn variation 
by spanwise wavelength, period and starting point in the base flow phase.

\section{Methodology}
\label{sec:method}
The flow over the blade is governed by the incompressible \NavSto\ 
equations, written in non-dimensional form as
\begin{subequations} 
\label{eq:fullnse} \begin{equation} \label{eq:nse} \partial_t\uvec = 
-(\uvec\bm{\cdot\nabla})\uvec
  - \bm{\nabla} p 
  + \Rey^{-1}\nabla^2\uvec \quad \hbox{in $\varOmega$}, 
\end{equation}
\begin{equation}
\label{eq:divcond}
\bm{\nabla \cdot} \uvec = 0 \quad\hbox{in $\varOmega$},
\end{equation}
\end{subequations}
where $\uvec(\xvec,t) = [u,v,w](x,y,z,t)$ is the velocity field,
$p(\xvec,t)$ is the kinematic (or modified) pressure field and
$\varOmega$ is the flow domain illustrated in
Figure~\ref{fig:mesh}. In what follows we define Reynolds number as
$\Rey = U_\infty c/\nu$, with $U_\infty$ being the inflow velocity
magnitude, $c$ the projection of the axial blade chord on the streamwise axis,
 and $\nu$ the kinematic
viscosity. 
Thus, we non-dimensionalise using $c$ as the length scale, $U_\infty$ as a velocity 
scale, so the time scale is then $c/U_\infty$.
 In the present work all numerical computations of the base
flows, whose two and three-dimensional energy growth characteristics
we are interested in, will exploit the homogeneity in $z$ and require
only a \twod\ computational domain.


We first consider a base flow about which we wish to study the linear 
stability.  The base flows for this problem are \twod, time-dependent flows 
that obey Equations \ref{eq:fullnse} with $\uvec=\Ubase$
and $P$ is defined as the associated base-flow pressure. The boundary 
conditions imposed on $\Omega$ in the base flow equations are 
uniform velocity $\U$ at the inflow, fully developed ($\partial_n 
\uvec=p=0$) at the outflow, periodic connectivity at the lower and upper 
boundaries and no-slip conditions at the blade surface. 


Our interest is in the evolution of infinitesimal perturbations
$\upert$ to the base flows.  The linearized \NavSto\ equations
governing these perturbations are found by substituting 
\begin{equation}
\uvec = \Ubase + \epsilon \upert\mbox{ \, and \, } p = \Pbase + \epsilon \ppert,
\label{eqn:Decomp2}
\end{equation}
 where $\ppert$ is the pressure perturbation, into the \NavSto\
equations and keeping the lowest order (linear) terms in
$\epsilon$. The resulting equations are
\begin{subequations}
\label{eq:pert_full}
\begin{equation}
\label{eq:pert_mom}
\partial_t\upert = 
-(\Ubase\bm{\cdot\nabla}) \upert
-(\upert\bm{\cdot\nabla}) \Ubase
  -\bm{\nabla} \ppert 
  + \Rey^{-1}\nabla^2 {\upert} \quad \hbox{in $\varOmega$}, 
\end{equation}
\begin{equation}
\label{eq:pert_div}
\bm{\nabla\cdot}{\upert} = 0 \quad\hbox{in $\varOmega$}.
\end{equation}
\end{subequations}

These equations are to be solved subject to appropriate initial 
conditions and the boundary conditions. The initial 
condition is an arbitrary incompressible flow which we denote by 
$\uvec_0$, i.e. $\upert(\xvec, t=t_0) = \uvec_0(\xvec)$.  The boundary 
conditions we consider are homogeneous Dirichlet on all 
boundaries, \ie $ \upert({\partial\varOmega},t) = \bm{0}$. As 
discussed in \cite{blbash08,bablsh08}, such homogeneous Dirichlet 
boundary conditions simplify the treatment of the adjoint problem 
because they lead to corresponding homogeneous Dirichlet boundary 
conditions on the adjoint fields.

We note that the action of Equations (\ref{eq:pert_full}) (a) and (b)
on an initial perturbation $\upert(\xvec,t_0)$ over time interval $\tau$
may be stated as
\begin{equation}
\label{eq.Aop}
\upert(\xvec,\tau) =\Aop(\tau)\upert(\xvec,t_0).
\end{equation}

\subsection{Linear asymptotic stability analysis}
\label{sec:method:lin_analysis}
The modal decomposition of this forward evolution
operator $\Aop(\tau)$ determines the asymptotic stability of the
base flow $\Ubase$. In this case the solution is proposed to be the
sum of eigenmodes,
\[\upert(\xvec,t)=\sum_j\exp(\lambda_jt)\umode_j(\xvec)+\text{c.c.},\]
and we obtain the eigenvalue problem
\begin{equation}
\Aop(\tau)\umode_j = \mu_j \umode_j, \quad \mu_j\equiv\exp(\lambda_j
\tau).
\label{eq.eigen}
\end{equation}

Since for the case of interest $\Ubase$ is $T$-periodic, we set 
$\tau=T$ and consider this as a temporal Floquet problem, in which 
case the $\mu_i$ are Floquet multipliers and the eigenmodes of 
$\Aop(\tau)$ are the $T$-periodic Floquet modes 
$\umode_j(\xvec,t+T)=\umode_j(\xvec,t)$ evaluated at a specific 
temporal phase.


\subsection{Optimal transient growth/ Singular value decomposition} 
\label{sec:method:dog}

Our primary interest is in the energy growth of perturbations over an 
arbitrary time interval, $\tau$. We treat $\tau$ and $t_0$ as parameters to 
be varied in this study.  As is conventional \cite{SchmidHenningson} we 
define transient growth with respect to the energy norm of the perturbation 
flow, derived from the $L_2$ inner product
\[ 2E(\upert)=(\upert,\upert) \equiv \int_\varOmega \upert\bm{\cdot}\upert~\cd V, \]
where $E$ is the kinetic energy per unit mass of a perturbation,
integrated over the full domain. The transient energy
growth over interval $\tau$ is
\begin{eqnarray*}
E(\tau)/E_0 &=& \left(\upert(\tau),\upert(\tau)\right) \\
&=& \left(\Aop(\tau)\upert(t_0),\Aop(\tau)\upert(t_0)\right)\\ 
& = & \left(\upert(t_0),\Aadj(\tau)\Aop(\tau)\upert(t_0)\right),
\end{eqnarray*}
where we introduce $\Aadj(\tau)$, the adjoint of the forward evolution
operator in \eqref{eq.Aop}. The action of $\Aadj(\tau)$ is obtained by
integrating the adjoint \LNS\ equations
\begin{subequations}
\label{eq.adj}
\begin{equation}
-\partial_t\uadj = 
-(\Ubase\bm{\cdot\nabla}) \uadj
+(\bm{\nabla}\Ubase)^\text{T}\bm{\cdot}\uadj
  -\bm{\nabla} \padj
  + \Rey^{-1}\nabla^2 {\uadj},
\end{equation}
\begin{equation}
\bm{\nabla\cdot}{\uadj} = 0 \quad\mbox{in $\varOmega$}
\end{equation}
\end{subequations}
backwards in time over interval $\tau$. The action of the symmetric component
operator $\Aadj(\tau)\Aop(\tau)$ on $\upert$ is obtained by serial
time integration of $\Aop(\tau)$ and $\Aadj(\tau)$, \ie we first use
$\upert(0)$ to initialise the integration of \eqref{eq:pert_full}
forwards in time over interval $\tau$, then use the outcome to
initialise the integration of \eqref{eq.adj} backwards in time over
the same interval.

The optimal perturbation is the eigenfunction of
$\Aadj(\tau)\Aop(\tau)$ corresponding to the compound operator's dominant
eigenvalue, and so we seek the dominant eigenvalues $\lambda_j$ and
eigenmodes $\bm{v}_j$ of the problem
\begin{equation*}
\Aadj(\tau)\Aop(\tau)\bm{v}_j = \lambda_j\bm{v}_j.
\end{equation*}

We use $G(\tau,t_0)$ to denote the maximum energy growth obtainable at time 
$\tau$ from initial time $t_0$, while the global maximum is denoted by 
${\Gmax}(t_0)=\max_{\tau} G(\tau,t_0)$.

Specifically,
\begin{equation}
G(\tau,t_0)=\max_j(\lambda_j) = \max_{\upert(0)}\frac{\left(\upert(\tau),\upert(\tau)\right)}{\left(\upert(t_0),\upert(t_0)\right)}
\end{equation}

We note that the
eigenfunctions $\bm{v}_j$ correspond to right singular vectors of
operator $\Aop(\tau)$, while their ($L_2$-normalised) outcomes
$\bm{u}_j$ under the action of $\Aop(\tau)$ are the left singular
vectors, \ie
\begin{equation}
\Aop(\tau)\bm{v}_j=\sigma_j\bm{u}_j,
\label{eq.svd}
\end{equation}
where the sets of vectors $\bm{u}_j$ and $\bm{v}_j$ are each
orthonormal. The singular values of $\Aop(\tau)$ are
$\sigma_j=\lambda_j^{1/2}$, where both $\sigma_j$ and $\lambda_j$ are
real and non-negative.

While long-time asymptotic growth is determined from the eigenvalue 
decomposition~\eqref{eq.eigen}, optimal transient growth is described 
in terms of the singular value decomposition~\eqref{eq.svd}. 
Specifically, the optimal initial condition and its (normalised) 
outcome after evolution over time $\tau$ are respectively the right 
and left singular vectors of the forward operator $\Aop(\tau)$ 
corresponding to the largest singular value. The square of that 
singular value is the largest eigenvalue of $\Aadj\Aop$ and is the 
optimal energy growth $G(\tau,t_0)$.

As already mentioned for an open flow, the most straightforward 
perturbation velocity boundary conditions to apply on both the inflow 
and outflow are homogeneous Dirichlet, \ie $\upert=0$, for both 
the forward and adjoint \LNS\ equations. The primitive variable, 
optimal growth formulation adopted in this work is discussed in 
further detail in \cite{blbash08,bablsh08} and follows almost directly 
from the treatments given by \cite{cobo00,luchini00,hbh05} for 
strictly parallel or weakly non-parallel basic states.


\subsection{Time integration and spatial discretisation}

\begin{figure}
\centering
\begin{minipage}[]{0.49\textwidth}
\psfrag{inflow}{inflow}
\psfrag{outflow}{outflow}
\psfrag{periodic}{periodic boundary}
\psfrag{boundary}{}
{\centering\includegraphics[bb=20 20 500 550,clip,width=\textwidth]
{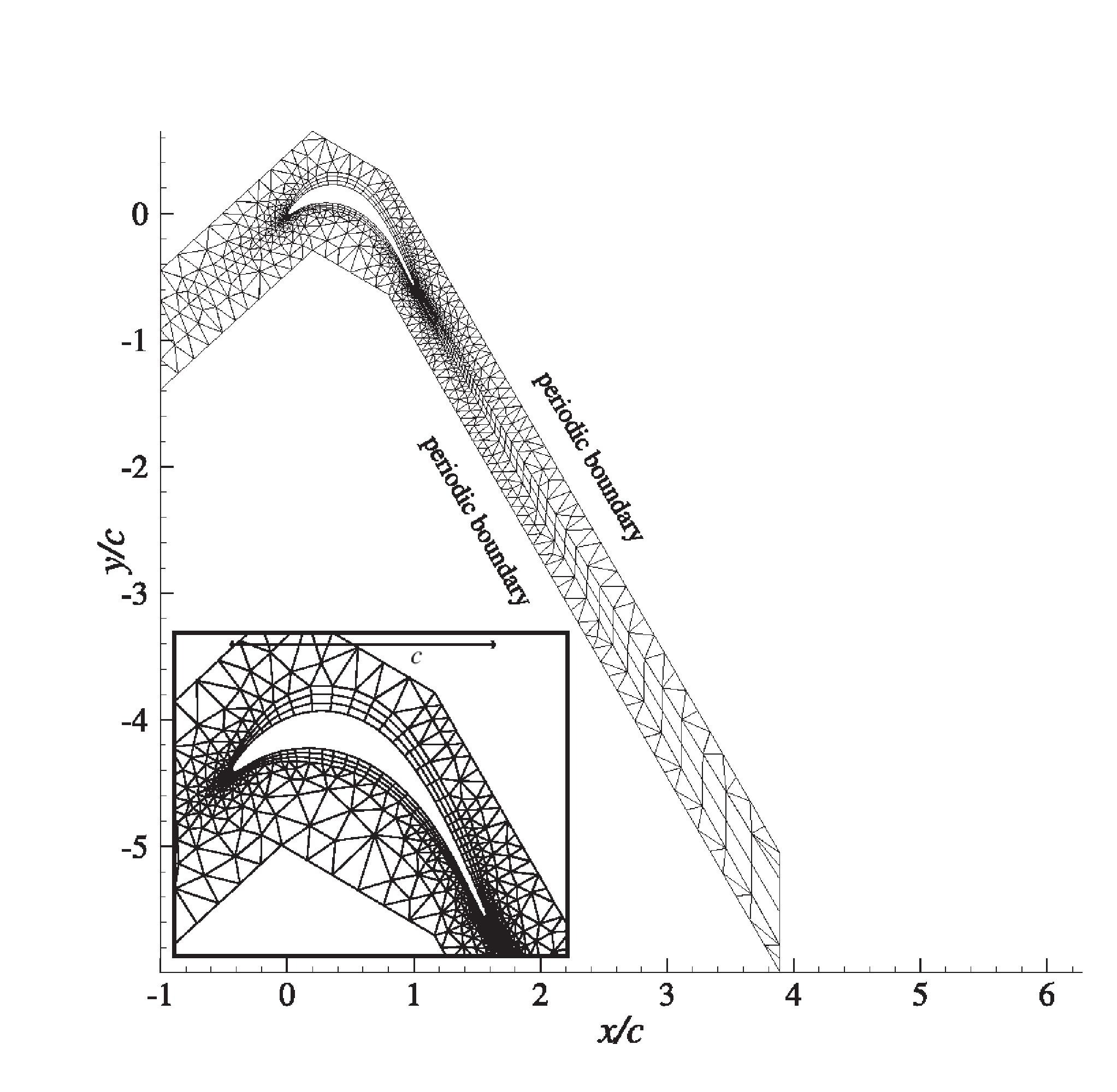}}
\end{minipage}
\begin{minipage}[]{0.49\textwidth}
\psfrag{x/c}{x/c}
\psfrag{y/c}{y/c}
\centering\includegraphics[bb=10 10 550 550,clip,width=\textwidth]{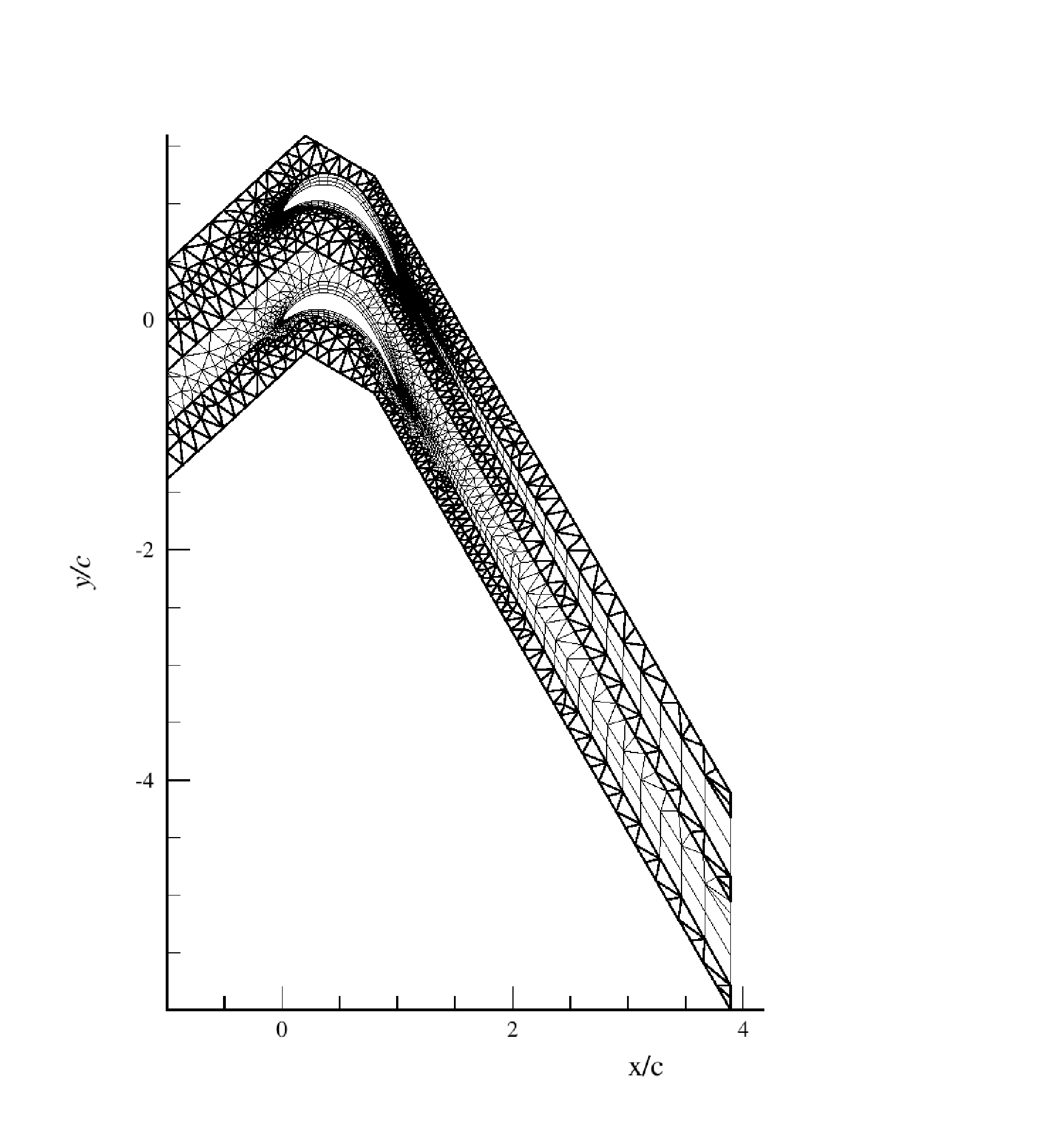}
\end{minipage}
\caption{Hybrid spectral/$hp$ macro element mesh made up of approximately 2000 
elements \emph{(left)} full mesh, \emph{(inset)} enlarged view around the 
blade and \emph{(right)} the double blade configuration. Within each element drawn, a polynomial expansion of order $p=6$ is applied}
\label{fig:mesh}
\end{figure}

Spectral/$hp$ elements \cite{KarniadakisSherwin} are used for spatial 
discretisation, coupled with a Fourier decomposition in the 
homogeneous direction. Time integration is carried out using a 
velocity-correction scheme \cite{Karniadakisetal,GuermondShen}.  The 
same discretisation and time integration schemes are used to compute 
base flows, and the actions of the forward and adjoint linearised \NS\ 
operators. The base flows are pre-computed and stored as data for the 
transient growth analysis in the form of $32$ time-slices. The base 
flow over one period of the evolution is reconstructed as required 
using Fourier interpolation.
As a check, the results for $\tau/T=4$ and $\beta=0$ were repeated with 16 time 
slices and found to differ by under $0.1\%$.

Figure \ref{fig:mesh} shows the computational domain for the T-106/300 low 
pressure turbine blade. The blade geometry is approximated by a cubic 
B-spline interpolation over 200 points to give a smooth flow surface. The 
hybrid mesh consists of approximately 2000 elements, 270 structured elements 
for the boundary layer around the blade surface and an unstructured mesh for 
the remainder of the field. The elements each have a polynomial order $p$ of 
$p=6$, with $(p+1)^2$ degrees of freedom for the quadrilateral elements and 
$(p+1)(p+2)/2$ for the triangular elements. Comparison with higher 
polynomial-order results showed that computations at the chosen polynomial 
order are sufficient to resolve the eigenvalue to about $0.2\%$.

\begin{figure}
\centering
\begin{minipage}[]{0.1\textwidth}
\vfill~\\
$t$\\ \vspace{1.87cm}\\
$t+ T/4$\\ \vspace{1.87cm}\\
$t+2T/4$\\ \vspace{1.87cm}\\
$t+ 3T/4$\\ \vspace{3cm}
\end{minipage}
\begin{minipage}[]{0.49\textwidth}
\resizebox{1.0\textwidth}{!}{\includegraphics{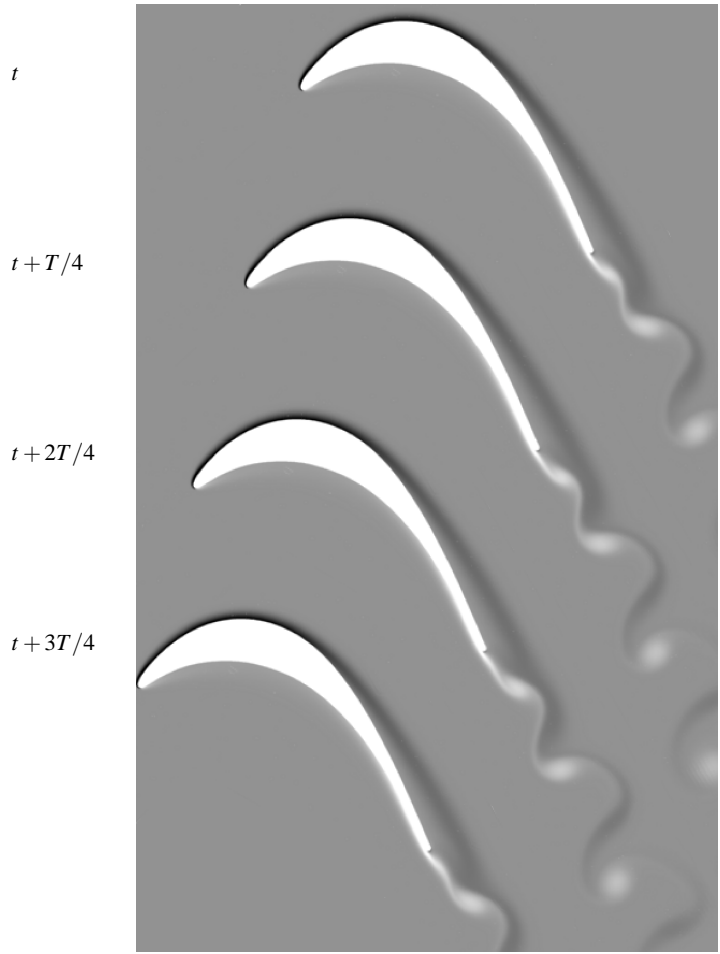}}
\end{minipage}
\caption{View of the periodic base flow at  \emph{(uppermost)} the beginning,
\emph{(second)} a quarter of the way through, \emph{(third)}
halfway through, \emph{(last)} three quarters of the way through one
base flow period. The figures show spanwise vorticity.}
\label{fig:baseflow}
\end{figure}

To calculate the base flow around which we apply perturbations, a 
two-dimensional DNS was performed. The two-dimensional periodic base flow is 
calculated at a Reynolds number $Re=2000$. The present study considers two 
and three-dimensional linear perturbations to this base flow.

Spatial periodicity between the planes has been assumed when computing the 
flow around one single blade. The imposition of zero Dirichlet boundary 
conditions on the outflow for all perturbation computations implies a 
natural limit to the integration period. Care was therefore be taken that 
the solution does not depend on the boundary conditions, in particular the 
outflow conditions. This was ensured by comparing results using an extended 
domain of twice the length to ensure independence of boundary conditions 
and solution for the cases under study. Doubling the domain length (for the 
$\beta=0$ case) changed the result for $\tau/T=4$ by less than $0.2\%$, for 
$\tau/T=8$ by less than $2\%$ and for $\tau/T=16$ by less than $8\%$. All 
these results are within plotting accuracy where presented. In the case of 
$T=16$, the results for the extended domain were presented.

\section{Behaviour of perturbations to a periodic base flow}
\label{sec:results}

Throughout the study the Reynolds number is fixed at 2000, corresponding to 
a periodic base flow (after the first Hopf bifurcation at Re=905). The 
shedding period of the base flow is $T=0.2676$ and the phase of the initial 
condition in the base flow shedding cycle is fixed at $t_0=0$ as depicted in 
Figure \ref{fig:baseflow} and as defined in the previous section, unless 
otherwise specified. As indicated in the introduction, the flow is best 
characterised as marginally unstable with Floquet multiplier ($\mu=1.00$).

\begin{figure}
	\centering
	\begin{minipage}{0.49\textwidth}\centering
	    \resizebox{1.0\textwidth}{!}{
%
%
\begin{psfrags}%
\psfragscanon%
%
\psfrag{s01}[t][t]{\color[rgb]{0,0,0}\setlength{\tabcolsep}{0pt}\begin{tabular}{c}$log_{10}(\tau/T)$\end{tabular}}%
\psfrag{s02}[b][b]{\color[rgb]{0,0,0}\setlength{\tabcolsep}{0pt}\begin{tabular}{c}$log_{10}(L_z/c)$\end{tabular}}%
\psfrag{s03}[][]{\color[rgb]{0,0,0}\setlength{\tabcolsep}{0pt}\begin{tabular}{c}1.5\end{tabular}}%
\psfrag{s04}[][]{\color[rgb]{0,0,0}\setlength{\tabcolsep}{0pt}\begin{tabular}{c}1.5\end{tabular}}%
\psfrag{s05}[][]{\color[rgb]{0,0,0}\setlength{\tabcolsep}{0pt}\begin{tabular}{c}1.5\end{tabular}}%
\psfrag{s06}[][]{\color[rgb]{0,0,0}\setlength{\tabcolsep}{0pt}\begin{tabular}{c}2\end{tabular}}%
\psfrag{s07}[][]{\color[rgb]{0,0,0}\setlength{\tabcolsep}{0pt}\begin{tabular}{c}2\end{tabular}}%
\psfrag{s08}[][]{\color[rgb]{0,0,0}\setlength{\tabcolsep}{0pt}\begin{tabular}{c}2\end{tabular}}%
\psfrag{s09}[][]{\color[rgb]{0,0,0}\setlength{\tabcolsep}{0pt}\begin{tabular}{c}2.5\end{tabular}}%
\psfrag{s10}[][]{\color[rgb]{0,0,0}\setlength{\tabcolsep}{0pt}\begin{tabular}{c}2.5\end{tabular}}%
\psfrag{s11}[][]{\color[rgb]{0,0,0}\setlength{\tabcolsep}{0pt}\begin{tabular}{c}3\end{tabular}}%
\psfrag{s12}[][]{\color[rgb]{0,0,0}\setlength{\tabcolsep}{0pt}\begin{tabular}{c}3\end{tabular}}%
\psfrag{s13}[][]{\color[rgb]{0,0,0}\setlength{\tabcolsep}{0pt}\begin{tabular}{c}3\end{tabular}}%
\psfrag{s14}[][]{\color[rgb]{0,0,0}\setlength{\tabcolsep}{0pt}\begin{tabular}{c}3.5\end{tabular}}%
\psfrag{s15}[][]{\color[rgb]{0,0,0}\setlength{\tabcolsep}{0pt}\begin{tabular}{c}3.5\end{tabular}}%
\psfrag{s16}[][]{\color[rgb]{0,0,0}\setlength{\tabcolsep}{0pt}\begin{tabular}{c}4\end{tabular}}%
\psfrag{s17}[][]{\color[rgb]{0,0,0}\setlength{\tabcolsep}{0pt}\begin{tabular}{c}4\end{tabular}}%
%
\psfrag{x01}[t][t]{-0.2}%
\psfrag{x02}[t][t]{0}%
\psfrag{x03}[t][t]{0.2}%
\psfrag{x04}[t][t]{0.4}%
\psfrag{x05}[t][t]{0.6}%
\psfrag{x06}[t][t]{0.8}%
\psfrag{x07}[t][t]{1}%
\psfrag{x08}[t][t]{1.2}%
%
\psfrag{v01}[r][r]{-0.5}%
\psfrag{v02}[r][r]{0}%
\psfrag{v03}[r][r]{0.5}%
\psfrag{v04}[r][r]{1}%
\psfrag{v05}[r][r]{1.5}%
\psfrag{v06}[r][r]{2}%
\psfrag{v07}[r][r]{2.5}%
\psfrag{v08}[r][r]{3}%
%
\resizebox{9cm}{!}{\includegraphics{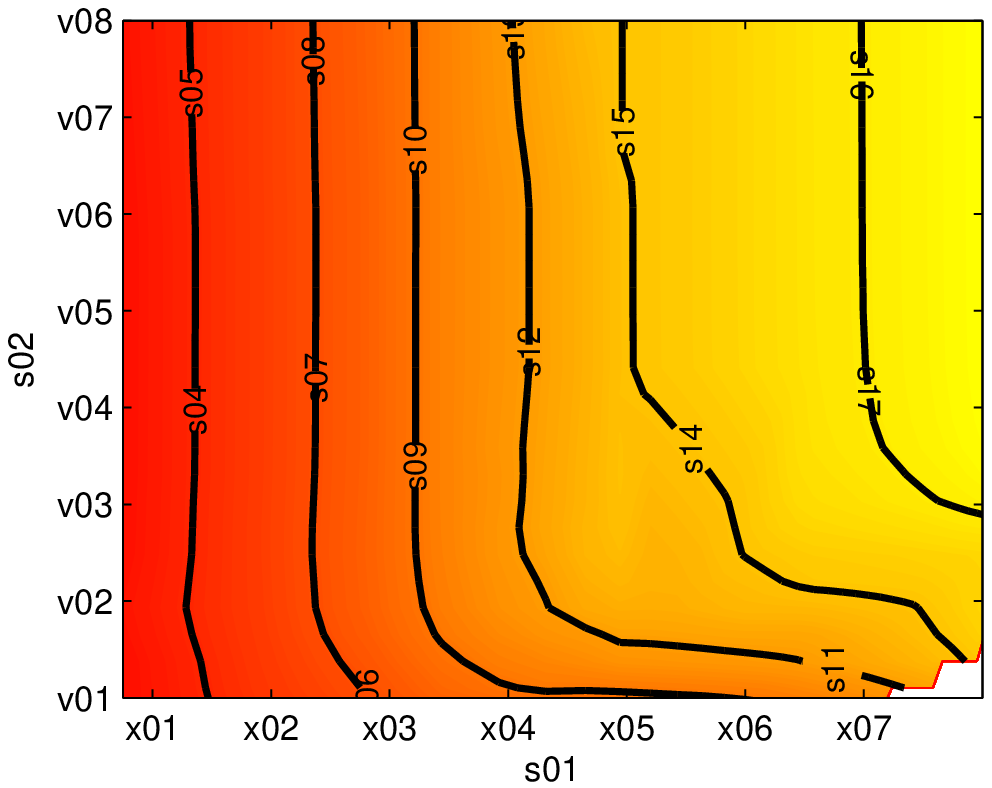}}%
\end{psfrags}%
%
	    	}
		\caption{Iso-curves showing transient growth in terms of $log_{10}(G)$ with varying $\tau$ and $L_z$, with fixed $t_0=0$.}
		\label{fig:T_Lz_surf}
    \end{minipage}
    \begin{minipage}{0.49\textwidth}\centering
        \resizebox{1.0\textwidth}{!}{
%
%
\begin{psfrags}%
\psfragscanon%
%
\psfrag{s01}[t][t]{\color[rgb]{0,0,0}\setlength{\tabcolsep}{0pt}\begin{tabular}{c}$log_{10}(\tau/T)$\end{tabular}}%
\psfrag{s02}[b][b]{\color[rgb]{0,0,0}\setlength{\tabcolsep}{0pt}\begin{tabular}{c}$\beta c$\end{tabular}}%
\psfrag{s03}[][]{\color[rgb]{0,0,0}\setlength{\tabcolsep}{0pt}\begin{tabular}{c}1.5\end{tabular}}%
\psfrag{s04}[][]{\color[rgb]{0,0,0}\setlength{\tabcolsep}{0pt}\begin{tabular}{c}1.5\end{tabular}}%
\psfrag{s05}[][]{\color[rgb]{0,0,0}\setlength{\tabcolsep}{0pt}\begin{tabular}{c}1.5\end{tabular}}%
\psfrag{s06}[][]{\color[rgb]{0,0,0}\setlength{\tabcolsep}{0pt}\begin{tabular}{c}2\end{tabular}}%
\psfrag{s07}[][]{\color[rgb]{0,0,0}\setlength{\tabcolsep}{0pt}\begin{tabular}{c}2\end{tabular}}%
\psfrag{s08}[][]{\color[rgb]{0,0,0}\setlength{\tabcolsep}{0pt}\begin{tabular}{c}2\end{tabular}}%
\psfrag{s09}[][]{\color[rgb]{0,0,0}\setlength{\tabcolsep}{0pt}\begin{tabular}{c}2.5\end{tabular}}%
\psfrag{s10}[][]{\color[rgb]{0,0,0}\setlength{\tabcolsep}{0pt}\begin{tabular}{c}2.5\end{tabular}}%
\psfrag{s11}[][]{\color[rgb]{0,0,0}\setlength{\tabcolsep}{0pt}\begin{tabular}{c}2.5\end{tabular}}%
\psfrag{s12}[][]{\color[rgb]{0,0,0}\setlength{\tabcolsep}{0pt}\begin{tabular}{c}3\end{tabular}}%
\psfrag{s13}[][]{\color[rgb]{0,0,0}\setlength{\tabcolsep}{0pt}\begin{tabular}{c}3\end{tabular}}%
\psfrag{s14}[][]{\color[rgb]{0,0,0}\setlength{\tabcolsep}{0pt}\begin{tabular}{c}3\end{tabular}}%
\psfrag{s15}[][]{\color[rgb]{0,0,0}\setlength{\tabcolsep}{0pt}\begin{tabular}{c}3.5\end{tabular}}%
\psfrag{s16}[][]{\color[rgb]{0,0,0}\setlength{\tabcolsep}{0pt}\begin{tabular}{c}3.5\end{tabular}}%
\psfrag{s17}[][]{\color[rgb]{0,0,0}\setlength{\tabcolsep}{0pt}\begin{tabular}{c}4\end{tabular}}%
%
\psfrag{x01}[t][t]{-0.2}%
\psfrag{x02}[t][t]{0}%
\psfrag{x03}[t][t]{0.2}%
\psfrag{x04}[t][t]{0.4}%
\psfrag{x05}[t][t]{0.6}%
\psfrag{x06}[t][t]{0.8}%
\psfrag{x07}[t][t]{1}%
\psfrag{x08}[t][t]{1.2}%
%
\psfrag{v01}[r][r]{0}%
\psfrag{v02}[r][r]{1}%
\psfrag{v03}[r][r]{2}%
\psfrag{v04}[r][r]{3}%
\psfrag{v05}[r][r]{4}%
%
\resizebox{9cm}{!}{\includegraphics{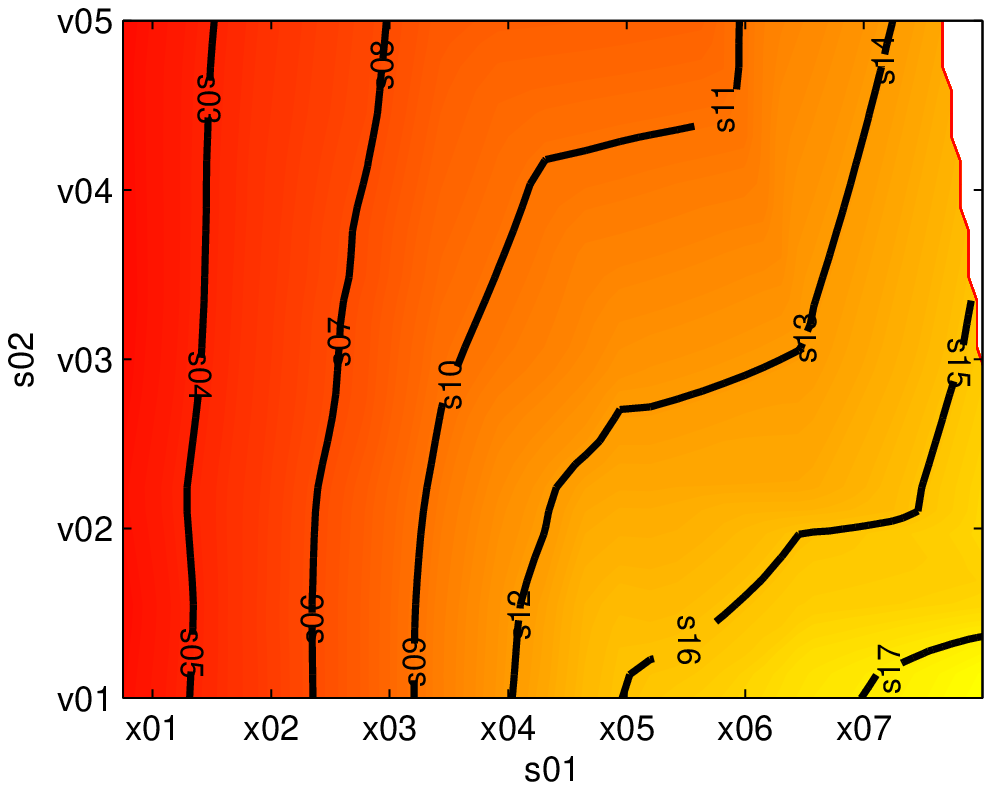}}%
\end{psfrags}%
%
        	}
		\caption{Iso-curves showing transient growth in terms of $log_{10}(G)$ with varying $\tau$ and $\beta$, with fixed $t_0=0$.}
		\label{fig:T_beta_surf}
	\end{minipage}
\end{figure}

The dependence of growth on $\tau$ and $L_z$ is shown in Figure 
\ref{fig:T_Lz_surf} and the same data is shown for $\beta = 2\pi/L_z$ in Figure 
 \ref{fig:T_beta_surf}. We find that, while for spanwise wavelengths 
above about $10c$, the growth achievable depends little on spanwise 
wavelength $L_z$, below $L_z \sim 10c$, the optimal growth in a given 
$\tau$ is limited, particularly for longer times. As $L_z$ tending to 
infinity represents a purely two-dimensional problem, it appears that 
it is the two-dimensional perturbations that dominate potential 
instabilities at this Reynolds number.  Growth of about over $10^4$ is 
achievable for $\tau\sim 10T$. For longer times the disturbance has 
convected too far downstream to be of interest.

To examine the case of shorter $\tau$ optimal disturbances we take 
$\tau=0.25T$ and vary $L_z$ as a parameter. Figure \ref{fig:Lz,T0.25,t00} 
shows the first two leading growth values associated to the two most 
significant optimum modes. It can be seen that the growth is moderate and 
concentrated at short $L_z$. The maximum occurs at $Lz=0.125c$ for this 
integration time. For slightly longer times ($\tau=T$ and above) the growth 
rises with $L_z$ and then flattens out, as shown in figure 
\ref{fig:Lz,T1.00,t00}. This effect is also evident in figure 
\ref{fig:Lz,T16.00,t00}.

\begin{figure}
	\centering
	\begin{minipage}{0.49\textwidth}\centering
		\resizebox{1.0\textwidth}{!}{
%
%
\begin{psfrags}%
\psfragscanon%
%
\psfrag{s01}[t][t]{\color[rgb]{0,0,0}\setlength{\tabcolsep}{0pt}\begin{tabular}{c}$L_z /c$\end{tabular}}%
\psfrag{s02}[b][b]{\color[rgb]{0,0,0}\setlength{\tabcolsep}{0pt}\begin{tabular}{c}$G_\tau$\end{tabular}}%
%
\psfrag{x01}[t][t]{$10^{-2}$}%
\psfrag{x02}[t][t]{$10^{0}$}%
\psfrag{x03}[t][t]{$10^{2}$}%
\psfrag{x04}[t][t]{$10^{4}$}%
%
\psfrag{v01}[r][r]{3}%
\psfrag{v02}[r][r]{4}%
\psfrag{v03}[r][r]{5}%
\psfrag{v04}[r][r]{6}%
\psfrag{v05}[r][r]{7}%
%
\resizebox{8cm}{!}{\includegraphics{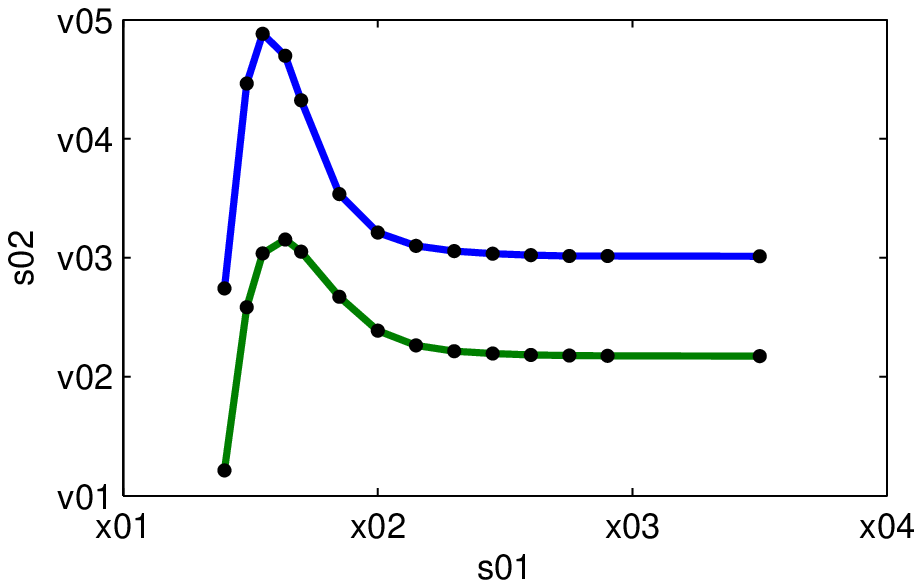}}%
\end{psfrags}%
%
			}
		\caption{Maximum growth as a function of $L_z$ for $\tau=0.25T$, $t_0=0$. The first and second singular values (squared) are shown.}
		\label{fig:Lz,T0.25,t00}
	\end{minipage}
	\begin{minipage}{0.49\textwidth}\centering
		\resizebox{1.0\textwidth}{!}{
%
%
\begin{psfrags}%
\psfragscanon%
%
\psfrag{s01}[t][t]{\color[rgb]{0,0,0}\setlength{\tabcolsep}{0pt}\begin{tabular}{c}$L_z/c$\end{tabular}}%
\psfrag{s02}[b][b]{\color[rgb]{0,0,0}\setlength{\tabcolsep}{0pt}\begin{tabular}{c}$G_\tau$\end{tabular}}%
%
\psfrag{x01}[t][t]{$10^{-1}$}%
\psfrag{x02}[t][t]{$10^{0}$}%
\psfrag{x03}[t][t]{$10^{1}$}%
\psfrag{x04}[t][t]{$10^{2}$}%
\psfrag{x05}[t][t]{$10^{3}$}%
%
\psfrag{v01}[r][r]{30}%
\psfrag{v02}[r][r]{40}%
\psfrag{v03}[r][r]{50}%
\psfrag{v04}[r][r]{60}%
\psfrag{v05}[r][r]{70}%
%
\resizebox{8cm}{!}{\includegraphics{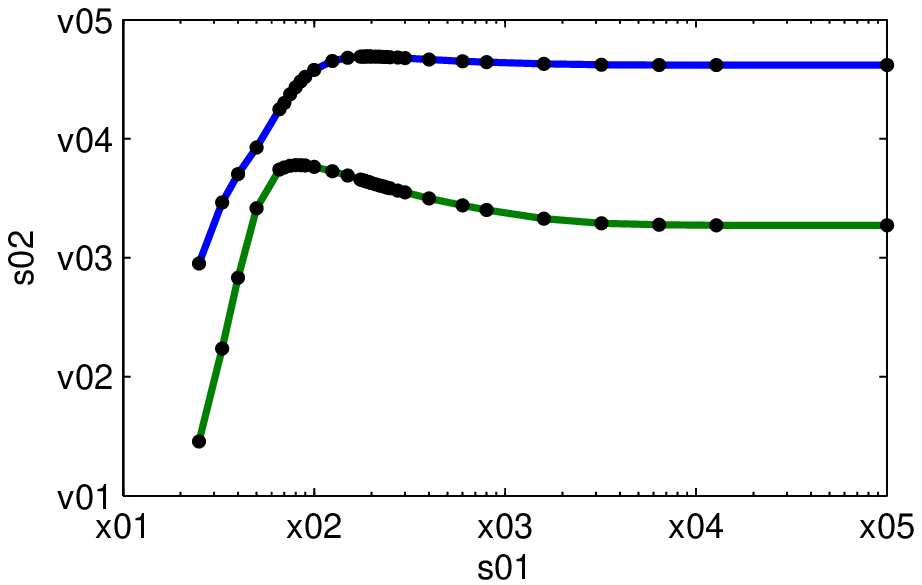}}%
\end{psfrags}%
%
			}
		\caption{Maximum growth as a function of $L_z$ for $\tau=T$, $t_0=0$}
		\label{fig:Lz,T1.00,t00}
	\end{minipage}
\end{figure}

\begin{figure}
	\centering
	\begin{minipage}{0.49\textwidth}\centering
		\resizebox{1.0\textwidth}{!}{
%
%
\begin{psfrags}%
\psfragscanon%
%
\psfrag{s01}[t][t]{\color[rgb]{0,0,0}\setlength{\tabcolsep}{0pt}\begin{tabular}{c}$L_z /c$\end{tabular}}%
\psfrag{s02}[b][b]{\color[rgb]{0,0,0}\setlength{\tabcolsep}{0pt}\begin{tabular}{c}$G_\tau$\end{tabular}}%
%
\psfrag{x01}[t][t]{$10^{-1}$}%
\psfrag{x02}[t][t]{$10^{0}$}%
\psfrag{x03}[t][t]{$10^{1}$}%
\psfrag{x04}[t][t]{$10^{2}$}%
\psfrag{x05}[t][t]{$10^{3}$}%
%
\psfrag{v01}[r][r]{0}%
\psfrag{v02}[r][r]{0.5}%
\psfrag{v03}[r][r]{1}%
\psfrag{v04}[r][r]{1.5}%
\psfrag{v05}[r][r]{2}%
\psfrag{ypower1}[Bl][Bl]{$\times 10^{4}$}%
%
\resizebox{8cm}{!}{\includegraphics{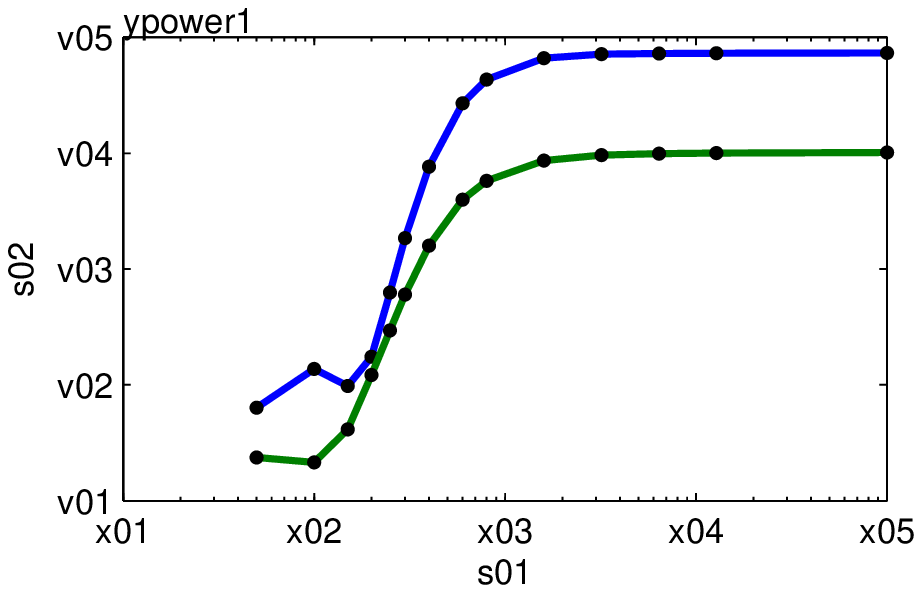}}%
\end{psfrags}%
%
			}
		\caption{Maximum growth as a function of $L_z$ for $\tau=16T$, $t_0=0$}
		\label{fig:Lz,T16.00,t00}
	\end{minipage}
	\begin{minipage}{0.49\textwidth}\centering
		\resizebox{1.0\textwidth}{!}{
%
%
\begin{psfrags}%
\psfragscanon%
%
\psfrag{s01}[t][t]{\color[rgb]{0,0,0}\setlength{\tabcolsep}{0pt}\begin{tabular}{c}$\tau/T$\end{tabular}}%
\psfrag{s02}[b][b]{\color[rgb]{0,0,0}\setlength{\tabcolsep}{0pt}\begin{tabular}{c}$G_\tau$\end{tabular}}%
%
\psfrag{x01}[t][t]{0}%
\psfrag{x02}[t][t]{5}%
\psfrag{x03}[t][t]{10}%
\psfrag{x04}[t][t]{15}%
\psfrag{x05}[t][t]{20}%
%
\psfrag{v01}[r][r]{$10^{0}$}%
\psfrag{v02}[r][r]{$10^{2}$}%
\psfrag{v03}[r][r]{$10^{4}$}%
%
\resizebox{8cm}{!}{\includegraphics{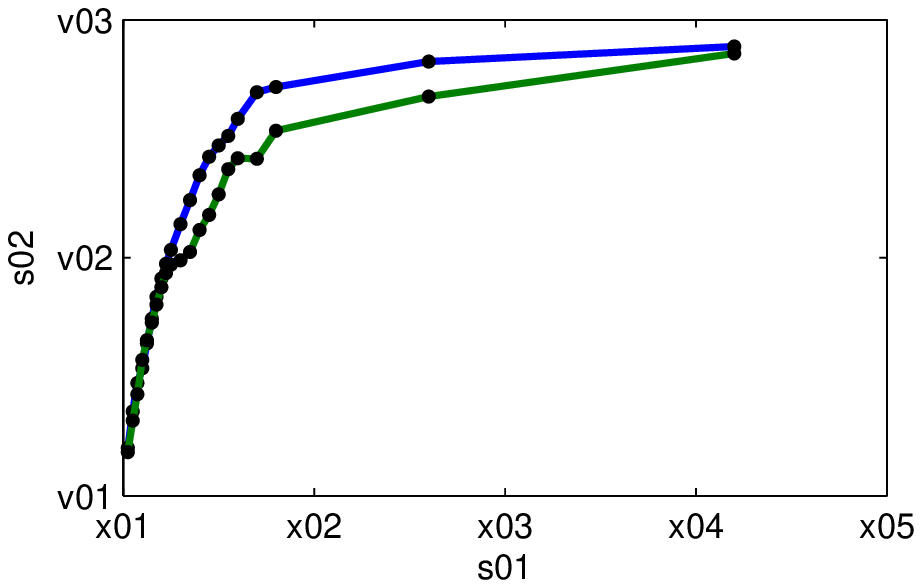}}%
\end{psfrags}%
%
		}
		\caption{Maximum growth as a function of $\tau$ for $L_z=1.95c$, $t_0=0$}
		\label{fig:T,Lz1.95,t00}
	\end{minipage}
\end{figure}

Taking the peak of the first mode at $\tau=T$ (see Figure 
\ref{fig:Lz,T1.00,t00}) and varying $\tau$ produces the results shown 
in Figure \ref{fig:T,Lz1.95,t00}. Again, strong growth is seen up to 
$\tau \sim 4T$ after which a plateauing is observed as the disturbance 
aligns to the least stable eigenmode (nearly marginally stable). The 
optimal mode for $\tau=T$ and $L_z=1.95c$ is shown in Figure 
\ref{fig:T01.00,LZ01.95,t00,s1}, which shows a disturbance beginning 
at the trailing edge and the shear layer exciting the near wake. The 
spanwise constant optimal mode for $\tau=8T$ is shown in Figure 
\ref{fig:T08.00,beta0,t00,s1}. In contrast to the short $\tau$ mode 
shown in the left panel of figure \ref{fig:T01.00,LZ01.95,t00,s1}, the 
wake mode is excited some distance downstream. Although the initial 
disturbance also involves separation at the shear layer and a 
disturbance at the trailing edge, the initial disturbance extends 
further up the blade surface. Figure \ref{fig:T01.00,LZ01.95,t00,3d} shows 
iso-surfaces of streamwise vorticity at $\omega_x=0.6$
of the same mode combined with $10\times$ the base flow.

\begin{figure}
\centering
\begin{minipage}[]{0.49\textwidth}
	\includegraphics[width=\textwidth]{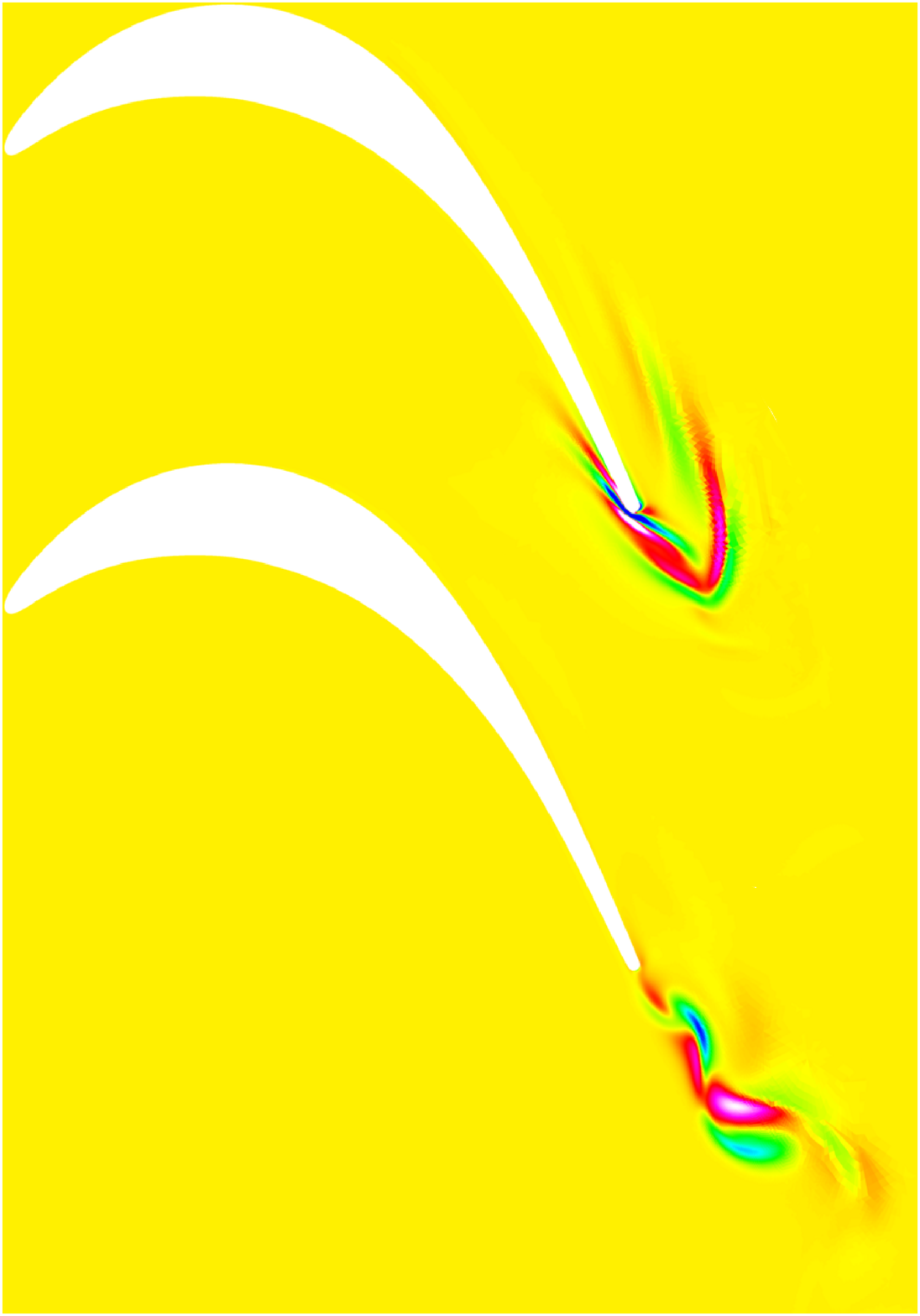}
\end{minipage}
\begin{minipage}[]{0.49\textwidth}
	\includegraphics[width=\textwidth]{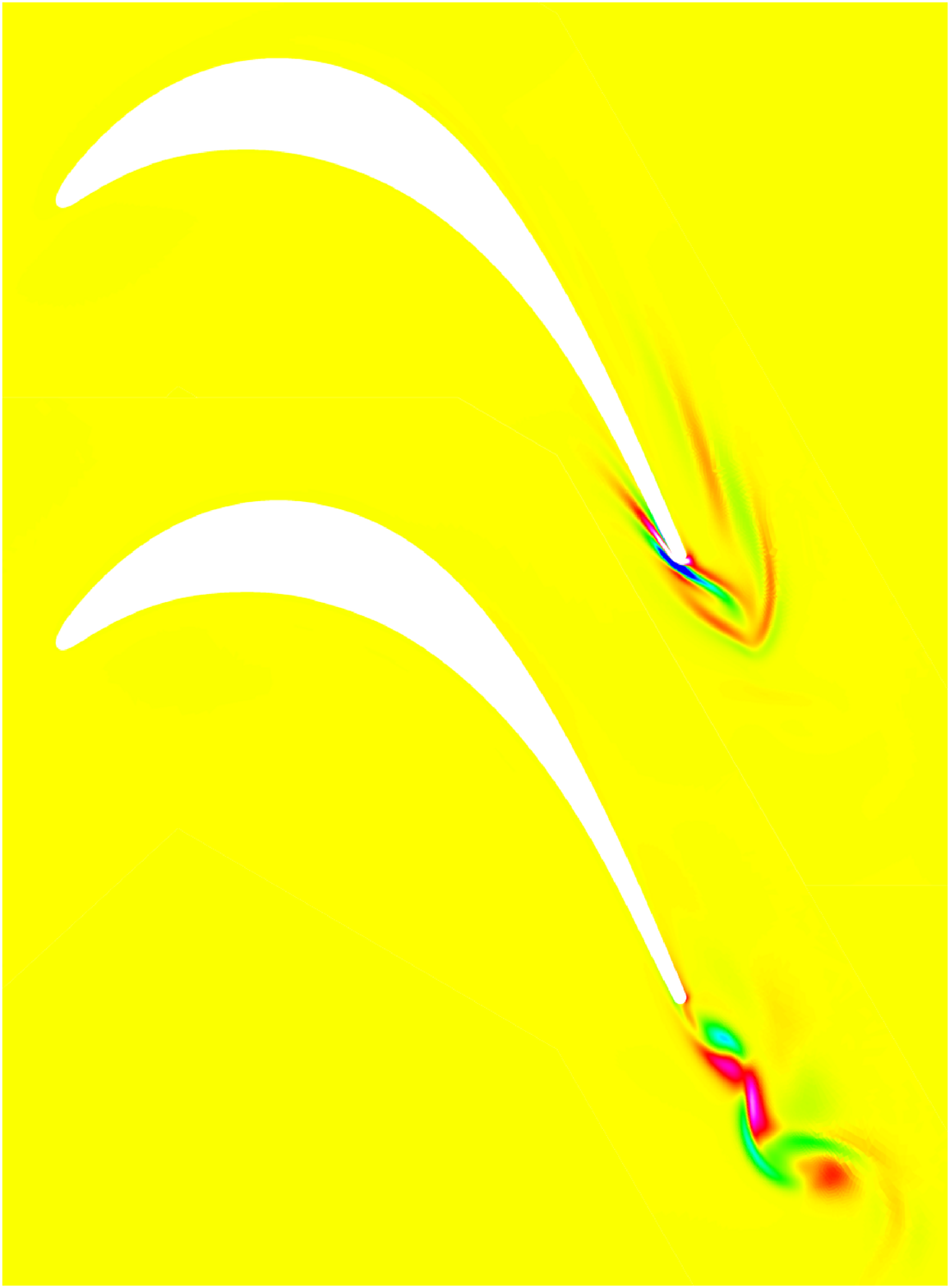}
\end{minipage}
\caption{Normalised leading mode initial condition \emph{(left upper)} 
and final state \emph{(left lower)} at $L_z=1.95c$, for $\tau=T$, at 
zero initial phase, showing spanwise vorticity. The right-hand plot 
shows the same for the mode associated with the second singular 
value.}
\label{fig:T01.00,LZ01.95,t00,s1}
\end{figure}

\begin{figure}
\centering
\begin{minipage}[]{0.49\textwidth}
	\includegraphics[bb=10 350 700 900,clip,width=\textwidth]{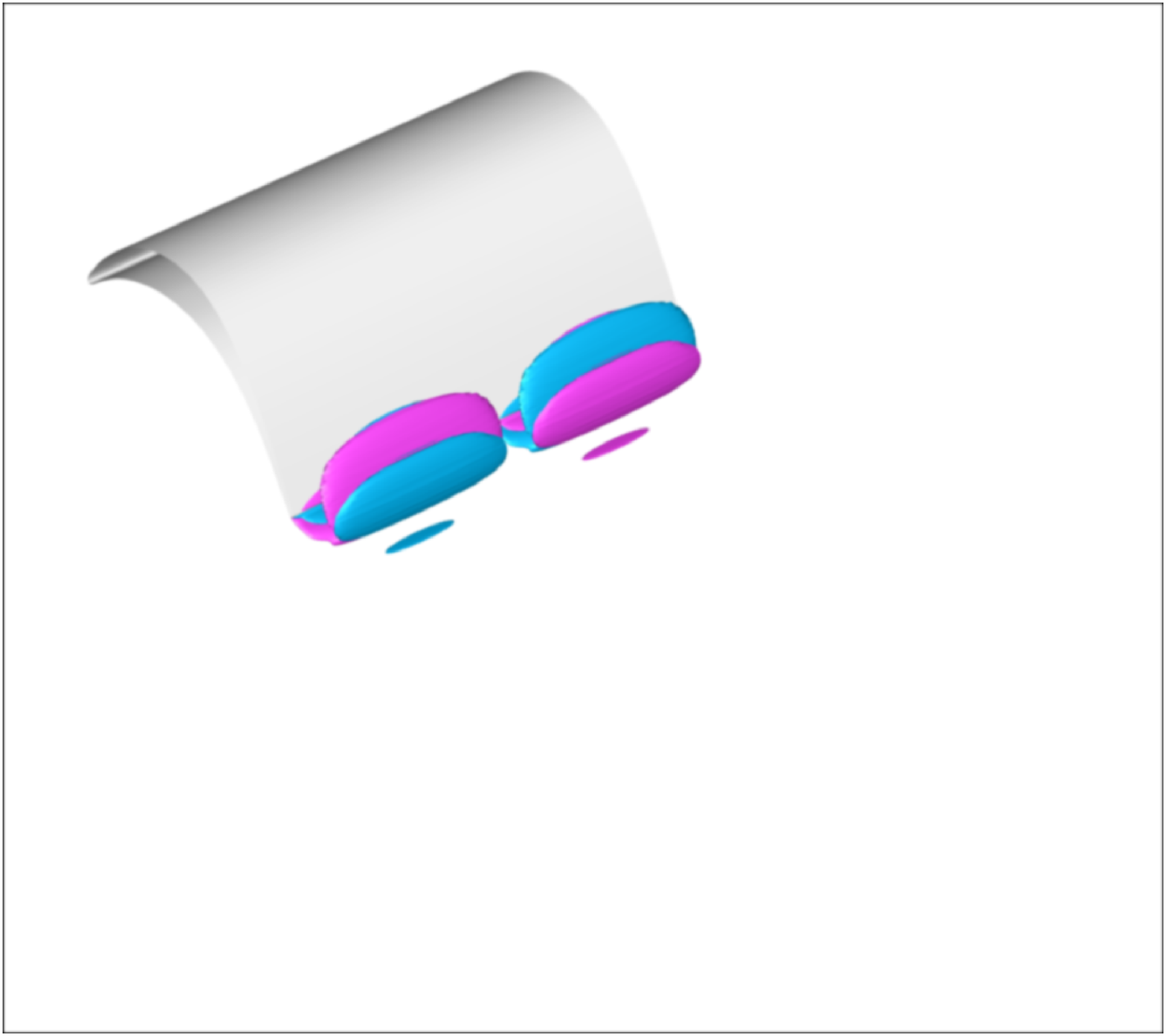}
\end{minipage}
\begin{minipage}[]{0.49\textwidth}
	\includegraphics[bb=10 350 700 900,clip,width=\textwidth]{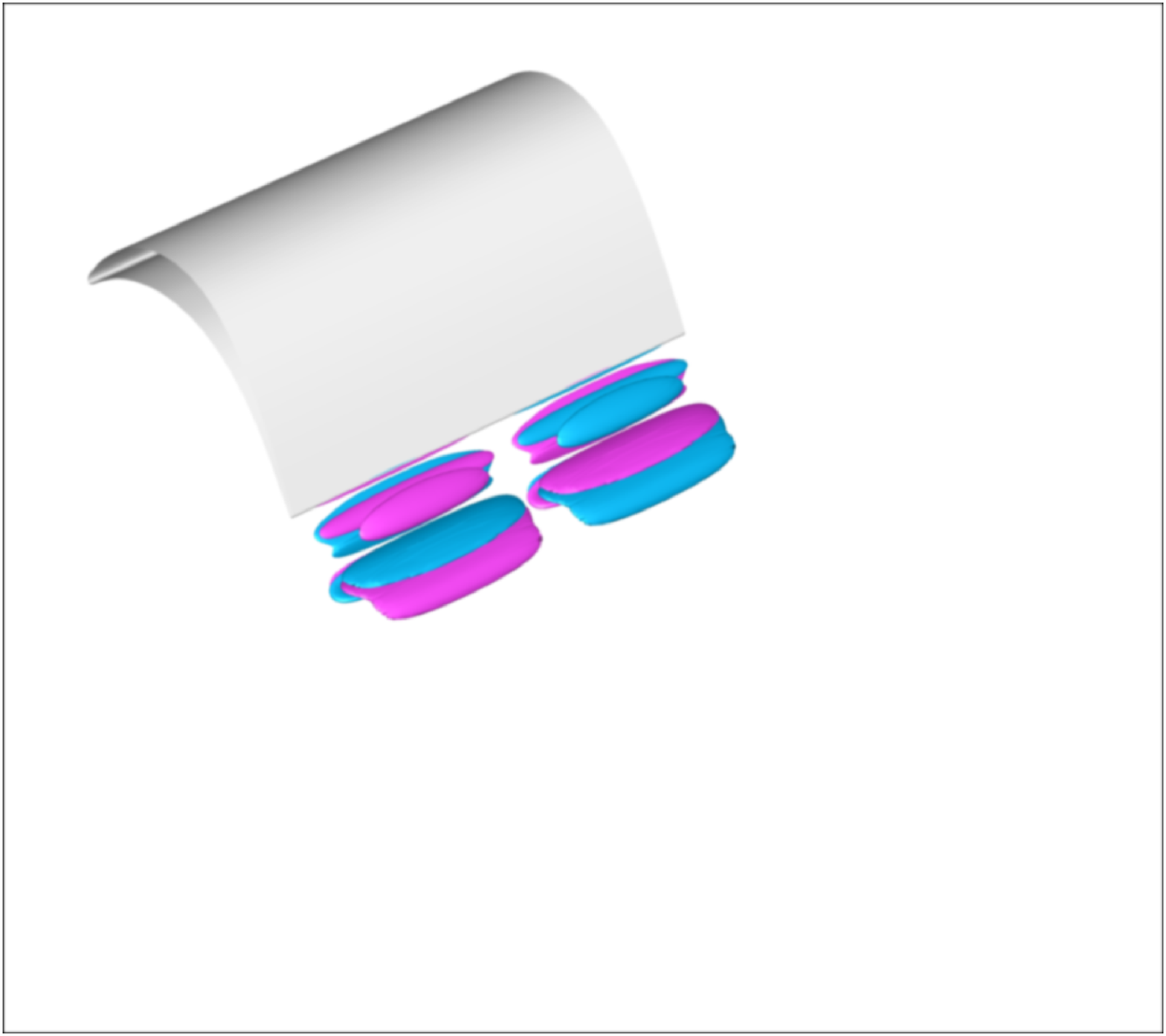}
\hspace{\textwidth}
\end{minipage}
\caption{Normalised leading mode initial condition \emph{(left)} 
and final state \emph{(right)} at $L_z=1.95c$, for $\tau=T$, at 
zero initial phase. The mode has been combined with $10\times$ the base flow
and the plots show iso-surfaces of streamwise vorticity at $\omega_x=\pm 0.6$}
\label{fig:T01.00,LZ01.95,t00,3d}
\end{figure}

\begin{figure}
\centering
\begin{minipage}[]{0.49\textwidth}
\includegraphics[width=\textwidth]{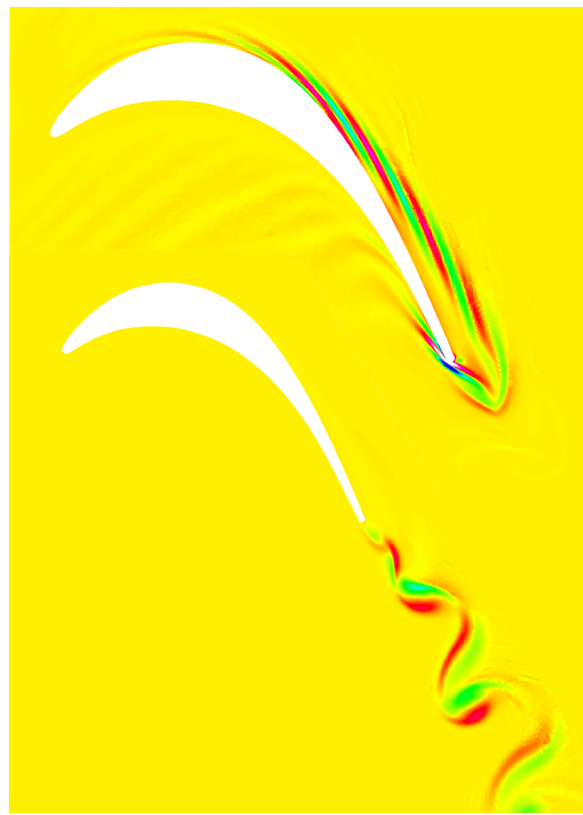}
\end{minipage}
\caption{Normalised leading mode initial condition \emph{(upper)} and final state \emph{(lower)}
at $\beta=0$ (2D case), for $\tau=8T$, at zero initial phase, showing spanwise vorticity}
\label{fig:T08.00,beta0,t00,s1}
\end{figure}

\begin{figure}
	\centering
	\begin{minipage}[]{0.49\textwidth}
	    \resizebox{1.0\textwidth}{!}{
%
%
\begin{psfrags}%
\psfragscanon%
%
\psfrag{s01}[b][b]{\color[rgb]{0,0,0}\setlength{\tabcolsep}{0pt}\begin{tabular}{c}$\beta c$\end{tabular}}%
\psfrag{s02}[t][t]{\color[rgb]{0,0,0}\setlength{\tabcolsep}{0pt}\begin{tabular}{c}$t_0/T$\end{tabular}}%
\psfrag{s03}[][]{\color[rgb]{0,0,0}\setlength{\tabcolsep}{0pt}\begin{tabular}{c}4.6\end{tabular}}%
\psfrag{s04}[][]{\color[rgb]{0,0,0}\setlength{\tabcolsep}{0pt}\begin{tabular}{c}4.6\end{tabular}}%
\psfrag{s05}[][]{\color[rgb]{0,0,0}\setlength{\tabcolsep}{0pt}\begin{tabular}{c}4.6\end{tabular}}%
\psfrag{s06}[][]{\color[rgb]{0,0,0}\setlength{\tabcolsep}{0pt}\begin{tabular}{c}4.6\end{tabular}}%
\psfrag{s07}[][]{\color[rgb]{0,0,0}\setlength{\tabcolsep}{0pt}\begin{tabular}{c}4.6\end{tabular}}%
\psfrag{s08}[][]{\color[rgb]{0,0,0}\setlength{\tabcolsep}{0pt}\begin{tabular}{c}4.6\end{tabular}}%
\psfrag{s09}[][]{\color[rgb]{0,0,0}\setlength{\tabcolsep}{0pt}\begin{tabular}{c}4.8\end{tabular}}%
\psfrag{s10}[][]{\color[rgb]{0,0,0}\setlength{\tabcolsep}{0pt}\begin{tabular}{c}4.8\end{tabular}}%
\psfrag{s11}[][]{\color[rgb]{0,0,0}\setlength{\tabcolsep}{0pt}\begin{tabular}{c}4.8\end{tabular}}%
\psfrag{s12}[][]{\color[rgb]{0,0,0}\setlength{\tabcolsep}{0pt}\begin{tabular}{c}4.8\end{tabular}}%
\psfrag{s13}[][]{\color[rgb]{0,0,0}\setlength{\tabcolsep}{0pt}\begin{tabular}{c}4.8\end{tabular}}%
\psfrag{s14}[][]{\color[rgb]{0,0,0}\setlength{\tabcolsep}{0pt}\begin{tabular}{c}4.8\end{tabular}}%
\psfrag{s15}[][]{\color[rgb]{0,0,0}\setlength{\tabcolsep}{0pt}\begin{tabular}{c}5\end{tabular}}%
\psfrag{s16}[][]{\color[rgb]{0,0,0}\setlength{\tabcolsep}{0pt}\begin{tabular}{c}5\end{tabular}}%
\psfrag{s17}[][]{\color[rgb]{0,0,0}\setlength{\tabcolsep}{0pt}\begin{tabular}{c}5\end{tabular}}%
\psfrag{s18}[][]{\color[rgb]{0,0,0}\setlength{\tabcolsep}{0pt}\begin{tabular}{c}5.2\end{tabular}}%
\psfrag{s19}[][]{\color[rgb]{0,0,0}\setlength{\tabcolsep}{0pt}\begin{tabular}{c}5.2\end{tabular}}%
\psfrag{s20}[][]{\color[rgb]{0,0,0}\setlength{\tabcolsep}{0pt}\begin{tabular}{c}5.2\end{tabular}}%
\psfrag{s21}[][]{\color[rgb]{0,0,0}\setlength{\tabcolsep}{0pt}\begin{tabular}{c}5.2\end{tabular}}%
\psfrag{s22}[][]{\color[rgb]{0,0,0}\setlength{\tabcolsep}{0pt}\begin{tabular}{c}5.2\end{tabular}}%
\psfrag{s23}[][]{\color[rgb]{0,0,0}\setlength{\tabcolsep}{0pt}\begin{tabular}{c}5.2\end{tabular}}%
\psfrag{s24}[][]{\color[rgb]{0,0,0}\setlength{\tabcolsep}{0pt}\begin{tabular}{c}5.4\end{tabular}}%
\psfrag{s25}[][]{\color[rgb]{0,0,0}\setlength{\tabcolsep}{0pt}\begin{tabular}{c}5.4\end{tabular}}%
\psfrag{s26}[][]{\color[rgb]{0,0,0}\setlength{\tabcolsep}{0pt}\begin{tabular}{c}5.4\end{tabular}}%
\psfrag{s27}[][]{\color[rgb]{0,0,0}\setlength{\tabcolsep}{0pt}\begin{tabular}{c}5.4\end{tabular}}%
\psfrag{s28}[][]{\color[rgb]{0,0,0}\setlength{\tabcolsep}{0pt}\begin{tabular}{c}5.4\end{tabular}}%
\psfrag{s29}[][]{\color[rgb]{0,0,0}\setlength{\tabcolsep}{0pt}\begin{tabular}{c}5.4\end{tabular}}%
%
\psfrag{x01}[t][t]{0}%
\psfrag{x02}[t][t]{0.2}%
\psfrag{x03}[t][t]{0.4}%
\psfrag{x04}[t][t]{0.6}%
\psfrag{x05}[t][t]{0.8}%
%
\psfrag{v01}[r][r]{0}%
\psfrag{v02}[r][r]{1}%
\psfrag{v03}[r][r]{2}%
\psfrag{v04}[r][r]{3}%
\psfrag{v05}[r][r]{4}%
%
\resizebox{9cm}{!}{\includegraphics{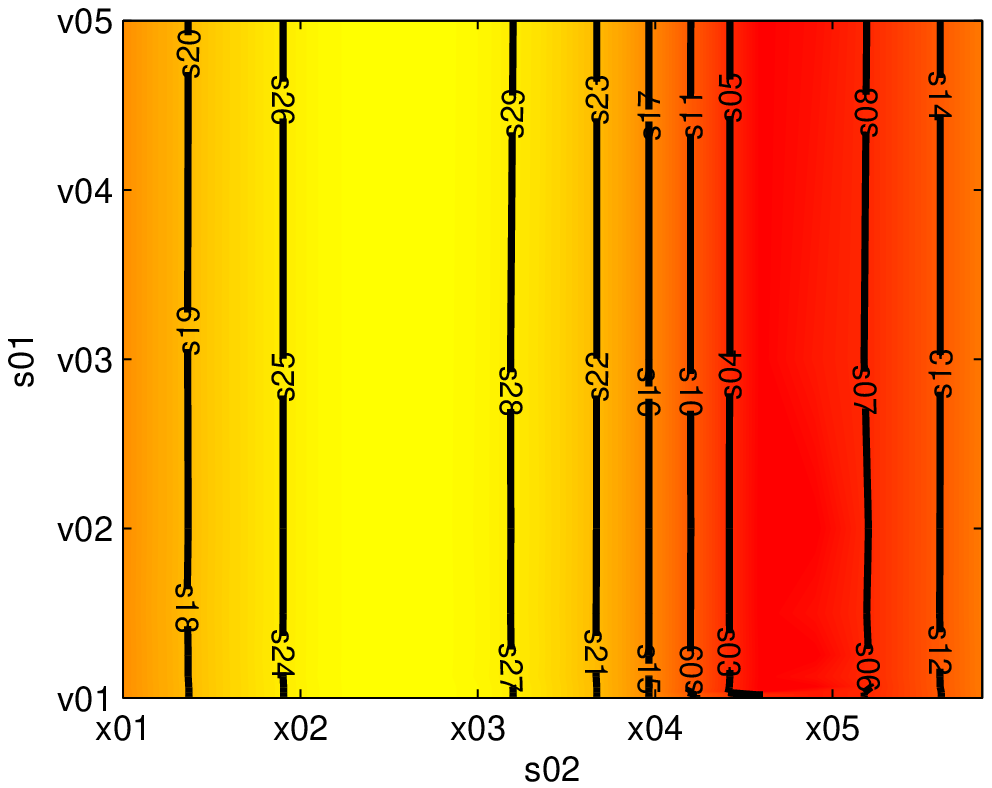}}%
\end{psfrags}%
%
	    	}
	\end{minipage}
	\begin{minipage}[]{0.49\textwidth}
	    \resizebox{1.0\textwidth}{!}{
%
%
\begin{psfrags}%
\psfragscanon%
%
\psfrag{s01}[b][b]{\color[rgb]{0,0,0}\setlength{\tabcolsep}{0pt}\begin{tabular}{c}$\tau/T$\end{tabular}}%
\psfrag{s02}[t][t]{\color[rgb]{0,0,0}\setlength{\tabcolsep}{0pt}\begin{tabular}{c}$t_0/T$\end{tabular}}%
\psfrag{s03}[][]{\color[rgb]{0,0,0}\setlength{\tabcolsep}{0pt}\begin{tabular}{c}1\end{tabular}}%
\psfrag{s04}[][]{\color[rgb]{0,0,0}\setlength{\tabcolsep}{0pt}\begin{tabular}{c}1\end{tabular}}%
\psfrag{s05}[][]{\color[rgb]{0,0,0}\setlength{\tabcolsep}{0pt}\begin{tabular}{c}1\end{tabular}}%
\psfrag{s06}[][]{\color[rgb]{0,0,0}\setlength{\tabcolsep}{0pt}\begin{tabular}{c}1\end{tabular}}%
\psfrag{s07}[][]{\color[rgb]{0,0,0}\setlength{\tabcolsep}{0pt}\begin{tabular}{c}1.5\end{tabular}}%
\psfrag{s08}[][]{\color[rgb]{0,0,0}\setlength{\tabcolsep}{0pt}\begin{tabular}{c}1.5\end{tabular}}%
\psfrag{s09}[][]{\color[rgb]{0,0,0}\setlength{\tabcolsep}{0pt}\begin{tabular}{c}1.5\end{tabular}}%
\psfrag{s10}[][]{\color[rgb]{0,0,0}\setlength{\tabcolsep}{0pt}\begin{tabular}{c}2\end{tabular}}%
\psfrag{s11}[][]{\color[rgb]{0,0,0}\setlength{\tabcolsep}{0pt}\begin{tabular}{c}2\end{tabular}}%
\psfrag{s12}[][]{\color[rgb]{0,0,0}\setlength{\tabcolsep}{0pt}\begin{tabular}{c}2\end{tabular}}%
\psfrag{s13}[][]{\color[rgb]{0,0,0}\setlength{\tabcolsep}{0pt}\begin{tabular}{c}2\end{tabular}}%
\psfrag{s14}[][]{\color[rgb]{0,0,0}\setlength{\tabcolsep}{0pt}\begin{tabular}{c}2.5\end{tabular}}%
\psfrag{s15}[][]{\color[rgb]{0,0,0}\setlength{\tabcolsep}{0pt}\begin{tabular}{c}2.5\end{tabular}}%
\psfrag{s16}[][]{\color[rgb]{0,0,0}\setlength{\tabcolsep}{0pt}\begin{tabular}{c}2.5\end{tabular}}%
\psfrag{s17}[][]{\color[rgb]{0,0,0}\setlength{\tabcolsep}{0pt}\begin{tabular}{c}3\end{tabular}}%
\psfrag{s18}[][]{\color[rgb]{0,0,0}\setlength{\tabcolsep}{0pt}\begin{tabular}{c}3\end{tabular}}%
\psfrag{s19}[][]{\color[rgb]{0,0,0}\setlength{\tabcolsep}{0pt}\begin{tabular}{c}3\end{tabular}}%
\psfrag{s20}[][]{\color[rgb]{0,0,0}\setlength{\tabcolsep}{0pt}\begin{tabular}{c}3\end{tabular}}%
\psfrag{s21}[][]{\color[rgb]{0,0,0}\setlength{\tabcolsep}{0pt}\begin{tabular}{c}3.5\end{tabular}}%
\psfrag{s22}[][]{\color[rgb]{0,0,0}\setlength{\tabcolsep}{0pt}\begin{tabular}{c}3.5\end{tabular}}%
\psfrag{s23}[][]{\color[rgb]{0,0,0}\setlength{\tabcolsep}{0pt}\begin{tabular}{c}3.5\end{tabular}}%
%
\psfrag{x01}[t][t]{0}%
\psfrag{x02}[t][t]{0.2}%
\psfrag{x03}[t][t]{0.4}%
\psfrag{x04}[t][t]{0.6}%
\psfrag{x05}[t][t]{0.8}%
%
\psfrag{v01}[r][r]{1}%
\psfrag{v02}[r][r]{2}%
\psfrag{v03}[r][r]{3}%
\psfrag{v04}[r][r]{4}%
\psfrag{v05}[r][r]{5}%
%
\resizebox{9cm}{!}{\includegraphics{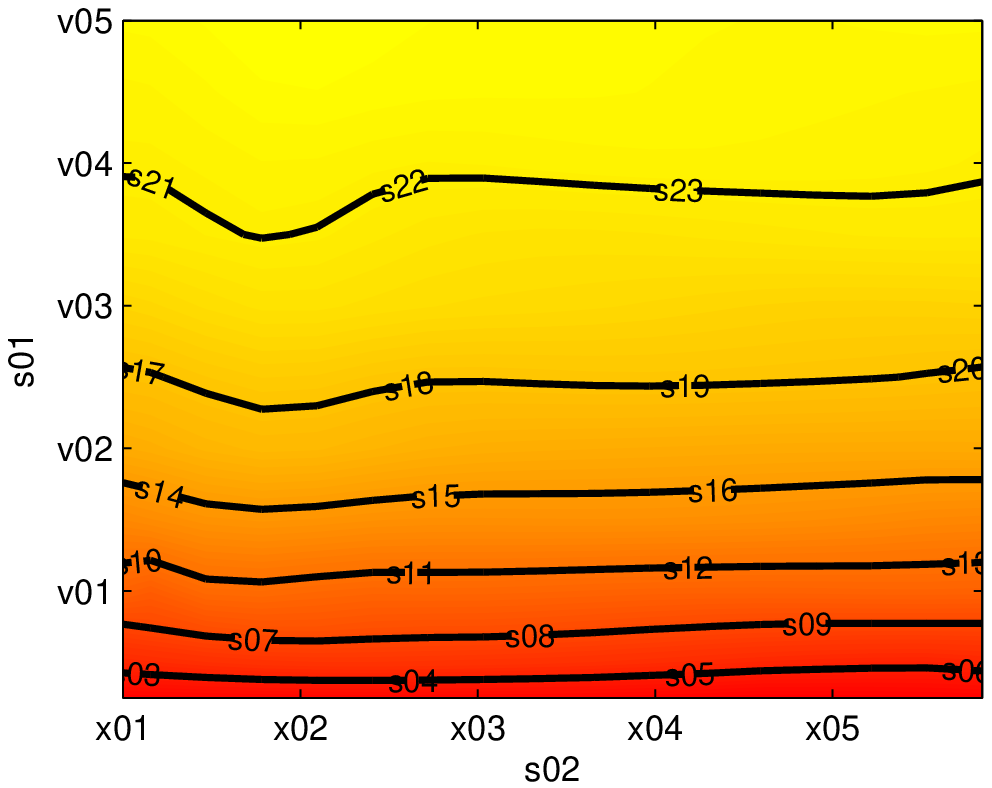}}%
\end{psfrags}%
%
	    }
	\end{minipage}
	\caption{\emph{(left)} $G_\tau$ with varying $t_0$ and $\beta$, $\tau=0.25T$;
	\emph{(right)} $log_{10}(G(\tau))$ with varying $\tau$ and $t_0$, $\beta=0$}
	\label{fig:t0_surfs}
\end{figure}

The preceding discussion has assumed an initial phase of $t_0=0$ 
(relating to the point in the shedding cycle) at $t=0$ as defined in 
figure \ref{fig:baseflow}. An investigation was also carried out to 
find the effects on the maximum attainable growth of varying this 
initial phase. At short growth horizons $\tau$, the growth achieved is 
only weakly dependent on $L_z$ and is achieved just before 
$t_0\sim0.2$ as can be seen in figure \ref{fig:t0_surfs}. Furthermore 
the figure also shows the dependence on initial phase remains weak 
over a range of integration times $\tau$. We conclude from the figure 
that the dependence of growth on the initial phase $t_0$ is relatively 
unimportant.

The relevant previous study \cite{Abdessemed09-1} demonstrated that 
although subharmonic effects were small, they were sufficient to 
induce asynchronous shedding and prevented the flow from going 
unstable. For the present problem we do not consider the asymptotic or 
long-time behaviour of perturbations but their behaviour over a 
shorter time horizon.

Results for the double-bladed mesh were tested for times up to eight 
shedding periods and agree well with the single blade situation. At 
these longer times, consideration of the least stable eigenmode would 
be more suitable. This leads to the conclusion that subharmonic 
effects are relatively unimportant in the transient growth problem and 
the assumed periodic boundary condition is a valid means to reduce the 
computational domain for the problem under study.

\section{Discussion and conclusion}
\label{sec:conclusion}
The transient behaviour of perturbations to linearised flow past a 
periodic array of T-106/300 low pressure turbine fan blade was 
investigated. The analysis was carried out at a Reynolds number of 
2000 associated with periodic vortex shedding, used as a periodic base 
flow.

It is known from asymptotic analysis \cite{Abdessemed09-1}, the flow 
past an array of these turbine blades is marginally stable. The 
current analysis shows that long wavelength optimal perturbations 
associated with long time integration periods convect far downstream 
and eventually align with the asymtotically least stable eigenmode. 
The discovery of converging optimum growth for long integration times 
confirms the lack of a strong asymptotic instability.

It is found that the long-wavelength perturbations tend toward a 
purely two-dimensional case and that these perturbations are 
associated to maximum optimum growth in the asymptotic case approached 
by long time-integration. However, as we already know from 
two-dimensional DNS, the two-dimensional baseflow is stable for 
nonlinear flows and therefore also stable in a linear sense. It 
therefore may be assumed that the identified optimum growth associated 
to long wavelength perturbations is less significant than 
perturbations that are limited in spanwise wavelength. When 
considering short integration times it is indeed found that 
short-wavelength perturbations are of higher significance.

Short integration times have maximum optimal growth at shorter 
spanwise wavelengths and convect only a short distance downstream, 
exciting the near wake. We might hypothesise that in the presence of 
the neglected nonlinearity, the associated optimum modes cause shear 
layer separation, triggering wake instability. This would have to be 
demonstrated in a full nonlinear DNS. Furthermore, the spanwise length 
of a real LPT blade is naturally limited, giving an additional reason 
why short wavelength perturbation are more important than long 
wavelength disturbances of theoretical spanwise extent.

Our results are consistent with the understanding that transient 
growth mechanisms are associated with shear in the base flow feeding 
the perturbation energy growth. The results presented are consistent 
therefore with the cylinder results reported in the literature (both 
on transient growth mechanisms \cite{Abdessemed09,shabshth06} and on 
receptivity via the adjoint \cite{Luchini07}) and with the previous 
work on the LPT fan blade used in this study \cite{Abdessemed09-1}.

It is hoped that the understanding developed will prove useful in 
controlling laminar boundary layer separation with a view to improving 
performance. For instance, the spatially periodic pattern in the shear 
layer can be quantified in the time-domain indicating disturbance 
frequencies susceptible to optimum amplification.

\begin{acknowledgements} A Sharma wishes to thank the UK Engineering and Physical 
Sciences Research Council (EPSRC) for their support and S Sherwin wishes to 
acknowledge financial support from the EPSRC Advanced Research 
Fellowship. Partial support has been received by the Air Force Office of Scientific
Research, under grant no. F49620-03-1-0295 to nu-modelling S.L.,
monitored by Dr T. Beutner (now at DARPA), Lt Col Dr R. Jefferies and
Dr J. D. Schmisseur of AFOSR and Dr S. Surampudi of the European
Office of Aerospace Research and Development.\end{acknowledgements}

\bibliographystyle{spmpsci}

\end{document}